\documentstyle[11pt,twoside]{article}
\begin{document}

\title{Magnetic fields and the large-scale structure}
\author{E. Battaner and E. Florido}
\maketitle

\begin{abstract}
The large-scale structure of the Universe has been observed to be
characterized by long filaments, forming polyhedra, with a remarkable
100-200 Mpc periodicity, suggesting a regular network. The
introduction of magnetic fields into the physics of the evolution of
structure formation provides some clues to understanding this unexpected
lattice structure. A relativistic treatment of the evolution of
pre-recombination inhomogeneities, including magnetic fields, is
presented to show that equivalent-to-present field strengths of the
order of $10^{-8}$ G could have played an important role. Primordial
magnetic tubes generated at inflation, at scales larger than the
horizon before recombination, could have produced filamentary density
structures, with comoving lengths larger than about 10 Mpc. Structures
shorter than this would have been destroyed by diffusion due to the
small pre-recombination conductivity. If filaments constitute a
lattice, the primordial magnetic field structures that produced the
post-recombination structures of matter, impose several restrictions on
the lattice. The simplest lattice compatible with these restrictions
is a network of octahedra contacting at their vertexes, which is
indeed identifiable in the observed distribution of superclusters.
\end{abstract}

The very large structure of the Universe is characterized by filaments
and voids. In supercluster distribution maps, such as those by
Tully et al. (1992) and Einasto et al. (1997), one can identify
filaments larger than 600 $h^{-1}Mpc$, such as the one connecting the
Tucana and the Ursa Major superclusters, probably extended to the
Draco supercluster. With galactic peculiar velocities of less than
$10^3 km/s$, in a time of the order of Hubble's, a galaxy is only able
to travel about 10 Mpc. Therefore it is difficult to explain
how the primordial mass inhomogeneities could be rearranged from a
chaotic distribution to one with ordered alignments. Galaxies, or
their pregalactic inhomogeneities, have no time to
redistribute themselves. Present hydrodynamic forces, and particularly magnetic
forces, have had no time to produce a redistribution and explain the
large  scale structures.

These large structures are also much larger than the horizon at
Recombination (about 10 Mpc, comoving length), which implies they
could have been originated only at the Inflation
epoch. Inflation magnetogenesis is one of the most interesting 
possibilities for the origin of
cosmic magnetic fields (Turner \& Widrow
1988; Ratra 1992; Garretson, Field, \& Carrol 1992; Dolgov 1993;
Gasperini \& Veneziano 1995 and others). If
present magnetic fields do not have an important influence
on the 100 Mpc large-scale structure, there remains the interesting
possibility that they were important in the past. In this paper, we
analyze the effects of primordial magnetic fields on the present
large-scale density distribution.

The history of magnetic fields during the different epochs of the
Universe is a complicated one, in which the Radiation Dominated Epoch was
critical, as the electron-photon interaction was responsible for a
resistivity capable of destroying small scale fields (Lesch \& Birk
1998). Magnetic diffusion induced by a
finite scalar conductivity cannot affect large scale
fields because it doesn't have sufficient time.  
It is likely that field structures larger than comoving 3 kpc were
able to survive and, with complete certainty, structures larger than
the horizon survived this hostile epoch.

Some of these field structures became sub-horizon after Recombination
and were not destroyed because, in this epoch, the assumption of
infinite conductivity is reasonably satisfied and magnetic
diffusivity is negligible at all scales.

As shown by Battaner, Florido \& Jimenez-Vicente (1997) and
Florido \& Battaner (1997), primordial magnetic flux tubes were
able to induce filamentary radiative energy density inhomogeneities
during the radiative epoch between Annihilation and Recombination, or
more precisely until the so-called Acoustic Epoch. Magnetic flux tubes
arise in cosmic MHD systems and are the
$\vec{B}$-structures necessary when magnetic coherence cells exist. They
anisotropically affect photon distribution because magnetic fields
are present in the energy-momentum tensor.

Energy density distributions created during this epoch produced
potential wells and seeds for baryonic and CDM inhomogeneities. They
originally consisted of filaments. Post recombination non-linear and
imperfect fluid effects distorted these structures, but the largest
ones were relatively unaffected and should be recognizable
today. Moreover, the original primordial magnetic flux tubes were distorted
by small scale effects, such as field amplification in growing
$\rho$-inhomogeneities, ejections by radio galaxies and other effects,
which again, kept the very large scales relatively
unaffected. Therefore, both the density and the magnetic
field large structures would have survived and should be recognizable
today.

A linear perturbation of the Maxwell, conservation of momentum-energy and
Einstein Field equations was carried out by Battaner, Florido \&
Jimenez-Vicente (1997) to study the evolution of magnetic field and
density inhomogeneities during the Radiation dominated era, even if
not all these equations are independent. The perturbed
Robertson-Walker metrics in a plane universe was considered.

A mean cosmological magnetic field cannot exist but a mean
magnetic energy density $<B^2/8\pi>$ may be non-vanishing. Provided 
that this is negligible
compared with the radiative energy density, the general laws of
expansion ($R \propto t^{1/2}$) and cooling ($T \propto R^{-1}$) are
unaffected by the presence of a magnetic energy.

During this epoch, the magnetic field strength always decreases, being
diluted by expansion, but the shape of the structures remains
unmodified in the expansion. The distribution of $\vec{B}_0 = \vec{B}
R^2$, where $R$ is the cosmological scale factor, taking its present
value as unity, is constant. This field, $\vec{B}_0$, would coincide
with the present field only if the 
 complicated post-recombination effects
had not distorted the structures and amplified the field.

The equivalent-to-present magnetic field strength during the
radiation dominated era should be of the order of  $10^{-8}
G$. If it were less than this, then magnetic fields would have no
influence on the structure. If higher, the growth of galaxies and
clusters would have proceeded too efficiently. A primordial magnetic flux
tube with $10^{-8}$ G at its centre would produce a density filament
with a relative over-density of $\delta = 5 \times 10^{-4}$ at
Recombination, starting with complete density homogeneity, or
considering primordial isocurvature inhomogeneities. After
Recombination $\delta$ has increased without the influence of magnetic
fields. These would have increased from $B_0 \sim 10^{-8} G$ before
Recombination to $B_0 \sim 1-3 \times 10^{-6}$ G at present, as
observed in clusters and in the intercluster medium (Kronberg, 1994). A
rough scheme is depicted in figure 1.

It is an observational fact that large scale filaments form polyhedra
which form a lattice (Broadhurst et al. 1990; Tully et al. 1992; Einasto
et al. 1997 and references therein). Battaner \& Florido (1997)
considered the properties of this lattice, as if they were produced by
primordial magnetic flux tubes, as present filaments would have
inherited the topological characteristics of the primordial
tubes. They concluded that the simplest network matching the magnetic
restrictions consists of octahedra only contacting at their vertexes.

If the real structures actually consisted of such octahedra, the
present supercluster distribution would remain as in an egg-carton
Universe. Battaner, Florido \& Garcia-Ruiz (1997) identified
octahedra only contacting at their vertexes in the real sky as being
all the
important superclusters and all the important voids (taken from the ETJEA,
i.e. Einasto et al. 1997, and the EETDA, i.e. Einasto et al. 1994
catalogues, respectively) forming part of the web. Therefore, 
the egg-carton network is
perfectly recognizable in the present large-scale structure. The size
of the octahedra would be $150 h^{-1} Mpc$.

As shown by Battaner (1998) this network is compatible with a fractal
structure, there being sub-octahedra within the octahedra, and so on. The lower
limit of the fractal range could be about 10 Mpc, because filaments
smaller than that probably had no chance of surviving the Radiation
Dominated era. The upper limit is that imposed by observations and
even by the present horizon. The fractal dimension would be either
1.77 or 2 depending on the ratio of octahedra/sub-octahedra sizes.

\vskip 0.5cm

\noindent {\bf{Conclusions}}

\vskip 0.5cm

Though some of the above ideas seems to be rather speculative, it is
very noticeable that the cosmic wave actually matches the theoretical
results. Our basic conclusion is that primordial magnetic fields have
played an important role in establishing the presently observed
supercluster network.

\vskip 0.5cm
                         
\noindent {\bf{References}}

\noindent Battaner, E. 1998, Astron. Astrophys., 334, 770

\noindent Battaner, E., Florido, E. \& Garcia-Ruiz, J.M. 1997,  Astron. Astrophys.,
327, 8

\noindent Battaner, E., Florido, E. \& Jimenez-Vicente, J. 1997,
 Astron. Astrophys., 326, 13

\noindent Broadhurst, T.J., Ellis, R.S., Koo, D.C. \& Szalay,
A.S. 1990, Nature, 343, 726

\noindent Dolgov, A.D. 1993, Phys. Rev. D, 48, 2499

\noindent Einasto, M., Einasto, J., Tago, E., Dalton, G.B. \&
Andernach, H. 1994, Mon. Not. Roy. Astron. Soc., 269, 301
 
\noindent Einasto, M., Tago, E., Jaaniste, J., Einasto, J. \&
Andernach, H. 1997, Astron. Astrophys. Supp. Ser., 123, 129

\noindent Florido, E. \& Battaner, E. 1997,  Astron. Astrophys., 327, 1

\noindent Garretson, W.D., Field, G.B. \& Carrol, S.M. 1992,
Phys. Rev D,
46, 5346

\noindent Gasperini, M. \& Veneziano, G. 1995, Astroparticle Physics,
1, 317

\noindent Kronberg, P.P. 1994, Reports on Progress in Physics, 57, 325

\noindent Lesch, H. \& Birk, G. 1998, Phys. Plasmas, 5, 2773

\noindent  Ratra, B. 1992, Astrophys. J., 391, L1

\noindent Tully, R.B., Scaramella, R., Vettolani, G. \& Zamorani,
G. 1992, Astrophys. J., 388, 9

\noindent Turner, M.S. \& Widrow, L.M. 1988, Physical Rev D, 37, 2743

\end{document}